\documentstyle[12pt,amssymb,amsfonts,amsbsy]{article}

\begin{document}
\author{Ilja Schmelzer\thanks
       {WIAS Berlin}}

\title{A Metric Theory of Gravity with Condensed Matter
Interpretation}

\begin{abstract}
\sloppypar 
We define a metric theory of gravity with preferred Newtonian frame
$(X^i(x),T(x))$ by

\[L = L_{GR}
    + \Xi g^{\mu\nu}\delta_{ij}X^i_{,\mu}X^j_{,\nu}
    - \Upsilon g^{\mu\nu}T_{,\mu}T_{,\nu}
\]

It allows a condensed matter interpretation which generalizes LET to
gravity.

The $\Xi$-term influences the age of the universe. $\Upsilon>0$ allows
to avoid big bang singularity and black hole horizon formation. This
solves the horizon problem without inflation. An atomic hypothesis
solves the ultraviolet problem by explicit regularization.  We give a
prediction for cutoff length.

\end{abstract}

\maketitle

\section{Introduction}

The theory we propose here is a metric theory of gravity with a
predefined Newtonian background frame.  Variables are the metric
tensor field $g_{\mu\nu}(x)$, matter fields $\psi^m(x)$, and the
Galilean coordinates $X^i(x), T(x)$ of the preferred frame.  Compared
with GR, the Lagrangian of the theory contains additional terms which
depend on these preferred coordinates:

\[L = R + \Lambda + L_{matter}(g_{\mu\nu},\psi^m)
    + \Xi g^{\mu\nu}\delta_{ij}X^i_{,\mu}X^j_{,\nu}
    - \Upsilon g^{\mu\nu}T_{,\mu}T_{,\nu}\]

$\Xi$ and $\Upsilon$ are additional ``cosmological constants'' which
have to be defined by observation, $\delta_{ij}$ is a predefined
Euclidean metric in the Newtonian background space.  For the preferred
coordinates we obtain the harmonic condition:

\[\Box X^i = \Box T = 0 \]

The additional terms in the Lagrangian compared with GR lead to
additional terms in the Einstein equations. This distinguishes the
theory from attempts to combine unmodified Einstein equations with
harmonic coordinates, as GR with harmonic gauge
(cf. \cite{Fock},\cite{Kuchar}) or theories where the preferred frame
remains hidden (cf. \cite{Logunov}, \cite{Schmelzer}).

The additional terms lead to observable effects.  The predefined
Newtonian frame explains the flatness of the universe. A positive
value of $\Xi $ increases the age of the universe.  A positive value of
$\Upsilon $ avoids the big bang singularity and leads to time-symmetric
solutions with a big crash before the big bang.  This solves the
horizon problem of relativistic cosmology without inflation theory.

Similar to ``Planck ether'' concepts \cite{Jegerlehner},\cite{Volovik}
ultraviolet quantization problems are solved by explicit, physical
regularization related based on an ``atomic hypothesis'' for the
condensed matter interpretation.  This hypothesis predicts a cutoff
different from Planck length, which seems to increase together with
the universe.

\section{Motivation}

To motivate the Lagrange density, it is sufficient to motivate the
harmonic equations for $X^i, T$. This special requirement about
coordinate dependence allows to justify the Lagrangian of our theory
in a similar way as the assumption of independence from $X^i, T$
justifies the Lagrangian of GR.

The theory allows a condensed matter interpretation, which explains
these harmonic equations as conservation laws for condensed matter.
But we do not have to rely on the condensed matter interpretation.  We
consider here a new axiom for quasi-classical quantum gravity and
EPR-realism as independent motivations for our theory.  But there is
also a sufficient number of other problems of GR which disappear in
our theory: local energy and momentum density for the gravitational
field, definition of vacuum state and Fock space in semi-classical
gravity, the problem of time and topological foam in quantum gravity
-- all these problems are closely related with the non-existence of a
Newtonian framework in GR which is available in our theory.

\subsection{Lagrange formalism}

If we require the harmonic condition as the equation for the preferred
coordinates, the general form of the Lagrangian is a simple consequence.

Indeed, once we handle the preferred coordinates as independent
fields, we can require covariance of the equations without restricting
generality.  Thus, we have to find a Lagrangian

\[ L = L(g_{\mu\nu},\psi^m,X^i,T) \]

for a covariant set of equation which contains the harmonic equations
$\Box X^i = \Box T = 0$. The simplest way to obtain covariant
equations is a covariant Lagrangian. The simplest way to obtain the
harmonic equations is to use standard scalar Lagrangians for $X^i, T$
and to assume that the remaining part does not depend on $X^i, T$.
But that means that the requirements for the remaining part are the
same as the standard requirements for a general-relativistic
Lagrangian -- thus, de-facto we have obtained our Lagrangian

\[L = L_{GR}(g_{\mu\nu},\psi)
    + \Xi g^{\mu\nu}\delta_{ij}X^i_{,\mu}X^j_{,\nu}
    - \Upsilon g^{\mu\nu}T_{,\mu}T_{,\nu}
\]

This is not a strong derivation -- we have preferred the simplest
possibilities instead of considering the general case.  But this seems
justified by Occam's razor, and is sufficient to explain the
Lagrangian.  Thus, to explain the equations of our theory completely
it is sufficient to explain the harmonic equation for $X^i,T$.

\subsection{Condensed Matter Interpretation}

The theory allows a reformulation in terms of condensed matter theory.
Instead of the gravitational field $g_{\mu\nu}$, we introduce
classical condensed matter variables -- density $\rho(x,t)$, velocity
$v^i(x,t)$, and stress tensor $\sigma^{ij}(x,t)$ -- by the following
formulas:

\begin{eqnarray} \label{gdef}
 \hat{g}^{00} = g^{00} \sqrt{-g} &=  &\rho \\
 \hat{g}^{i0} = g^{i0} \sqrt{-g} &=  &\rho v^i \\
 \hat{g}^{ij} = g^{ij} \sqrt{-g} &=  &\rho v^i v^j - \sigma^{ij}
\end{eqnarray}

These condensed matter variables are Galilean covariant.  If
$\rho(x,t) > 0$ and $\sigma^{ij}(x,t)$ is positive definite they
define a Lorentz metric.  Moreover, the harmonic equation transforms
into classical conservation laws:

\begin{eqnarray} \label{conservationlaw}
\partial_t \rho + \partial_i (\rho v^i) &= &0 \\
\partial_t (\rho v^j) + \partial_i(\rho v^i v^j - \sigma^{ij}) &= &0
\end{eqnarray}

Note also the very natural expression for the additional terms of the
Lagrangian:

\[ L\sqrt{-g}= L_{GR}\sqrt{-g}+\Xi (\rho v^2-\sigma^{ii})-\Upsilon \rho\]

Note that the conservation laws remain unchanged even if there are
other ``matter fields'' $\psi^m(x)$.  That means, these fields do not
describe external matter, but inner steps of freedom of the condensed
matter itself.  Thus, the condensed matter is described by $\rho, v^i,
\sigma^{ij}$ and ``inner steps of freedom'' $\psi^m$.  That's why the
momentum related with inner steps of freedom is already taken into
account.  

In some sense, this interpretation of ``matter fields'' in this
condensed matter interpretation unifies gravity with usual matter
fields.  More important is that it explains the harmonic equations for
$X^i, T$ in the presence of matter fields, and therefore the whole
theory.

The non-gravity limit of the condensed matter interpretation is
Lorentz ether theory.  Thus, this interpretation may be considered as
a generalization of Lorentz ether theory to gravity.  This suggests to
name this interpretation {\it general ether theory}.

\subsection{Quantum gravity motivation}

It is straightforward that the introduction of a Newtonian framework
solves the most serious conceptual problems of GR quantization: the
problem of time \cite{Isham}, topological foam, the information loss
problem.  The problems with local energy and momentum density of the
gravitational field and the uncertainty of the definition of Fock
space and vacuum state in semi-classical QFT, which are also connected
with the absence of a preferred frame in GR, may be mentioned too.
But to introduce a Newtonian framework to solve these problems is
often criticized as an ad-hoc simplification.  That's why I prefer to
present a quantum gravity motivation of different type, related with
quasi-classical quantum gravity (superpositions of semi-classical
solutions).

If we consider superpositions of gravitational fields $g_{\mu\nu}(x)$,
the classical notion of covariance may be generalized in two ways:
c-covariance denotes covariance if we use the same diffeomorphism for
all fields, q-covariance allows different diffeomorphisms for
different fields \cite{Anandan}.  The Einstein equations are
q-covariant.  Canonical GR quantization also defines a q-covariant
theory.  Instead, our theory is c-covariant. To motivate our theory it
would be sufficient to motivate the existence of c-covariant objects.

For this purpose, let's consider the probability that a
super-positional state $|g^1\rangle+|g^2\rangle$ of gravitational
fields switches into $|g^1\rangle-|g^2\rangle$ because of
gravitational interaction with a test particle $\varphi$.  Let's
consider non-relativistic quantum gravity -- two-particle
Schr\"odinger theory with Newtonian potential, with particles g and
$\varphi$.  Here gravitational interaction transforms the initial
state into the state $|g^1,\varphi^1\rangle+|g^2,\varphi^2\rangle$.
We ignore the test particle and compute the resulting one-particle
state for g, which is in general a mixed state.  The transition
probability $|g^1\rangle+|g^2\rangle\to|g^1\rangle-|g^2\rangle$ is
${1\over2}(1-\langle\varphi^1|\varphi^2\rangle)$, thus, depends on the
scalar product $\langle\varphi^1|\varphi^2\rangle$.  

The natural generalization to semi-classical theory for
$\varphi^{1/2}(x)$ is the solution for the test particle on the
background $g^{1/2}_{\mu\nu}(x)$ created by the particle $g^{1/2}$.
The scalar product between these functions is only c-covariant, not
q-covariant.

As a new axiom for quantum gravity we propose that this scalar product
is well-defined and observable.  This axiom does not have the fault of
being an ad-hoc simplification to avoid topological problems, but is a
natural generalization of an observable of pre-relativistic quantum
gravity.  Thus, it is sufficiently motivated.  Moreover, it has some
beauty: it is a typical quantum observable with global character, it
does not depend on questionable assumptions about local measurements.

Once we accept scalar products, a preferred system of coordinates is a
very natural object.  It is natural to assume that the scalar products
define an isomorphism between the related $L^2$-spaces.  Such an
isomorphism allows to transfer the projective measure related with
position measurement on a simple fixed state (the ``vacuum'') to other
gravitational fields, thus, to define common coordinates on all
gravitational fields, with a common topology as a consequence.

Independent of the last argument, the axiom requires to reject
canonical GR quantization because of its q-covariance, while canonical
quantization of our theory allows to make c-covariant predictions for
such scalar products.

\subsection{Realistic motivation}

As shown by the proof of Bell's inequality \cite{Bell} and their
experimental falsification by Aspect \cite{Aspect} there is a
contradiction between Einstein causality and the EPR criterion of
reality \cite{EPR}.  This is usually interpreted as an experimental
falsification of EPR-realism.  But this is incorrect -- only if we
accept Einstein causality as an axiom, Aspect's experiment falsifies
EPR-realism.

We can as well turn the argument against Einstein causality.  We
simply use EPR-realism and causality as axioms.  With these axioms,
Aspect's experiment falsifies Einstein causality and allows to prove
the existence of a preferred foliation.  Moreover, the existence of a
preferred foliation allows to use Bohmian mechanics \cite{Bohm}
instead of quantum theory.  

The condensed matter interpretation of our theory defines such a
preferred foliation.

\section{Predictions}

For a metric theory of gravity with Einstein equations in the limit
$\Xi ,\Upsilon \to 0$ it is not problematic to fit existing
observation as well as GR fits observation.  Instead, it is a
non-trivial problem to distinguish the theory from GR by observation.
Nonetheless, especially if $\Upsilon>0$, this seems possible.

\subsection{$\Xi $ as a dark matter candidate}

Let's consider at first the homogeneous universe solutions of the
theory.  Because of the Newtonian background frame, only a flat
universe may be homogeneous. Thus, we make the ansatz:

\[ ds^2 = d\tau^2 - a^2(\tau)(dx^2+dy^2+dz^2) \]

Note that in this ansatz the universe does not really expand, the
observable expansion is an effect of shrinking rulers.  Below we
nonetheless use standard relativistic language. Using some matter with
$p = k \varepsilon$ we obtain the equations ($8\pi G = c = 1$):

\begin{eqnarray*}
3(\dot{a}/a)^2  &=& 
	- \Upsilon/a^6 + 3 \Xi/a^2 + \Lambda + \varepsilon\\
2(\ddot{a}/a) + (\dot{a}/a)^2 &=& 
	+ \Upsilon/a^6 +   \Xi/a^2 + \Lambda - k \varepsilon
\end{eqnarray*}

The influence of the $\Xi $-term on the age of the universe is easy to
understand.  For $\Xi >0$ it behaves like homogeneously distributed
dark matter with $p=-(1/3)\varepsilon$.  It influences the age of
the universe.  A similar influence on the age of the universe has a
non-zero curvature in GR cosmology.  It seems not unreasonable to hope
that a non-zero value for $\Xi $ may be part of the solution of the
dark matter problem.

\subsection{$\Upsilon >0$ solves the horizon problem without inflation}

Instead, $\Upsilon$ influences the early universe, its influence on
later universe may be ignored.  But, if we assume $\Upsilon >0$, the
qualitative behaviour of the early universe changes in a remarkable
way.  We obtain a lower bound $a_0$ for $a(\tau)$ defined by

\[ \Upsilon/a_0^6 = 3 \Xi/a_0^2 + \Lambda + \varepsilon \]

The solution becomes symmetric in time, with a big crash followed by
a big bang.  For example, if $\varepsilon = \Xi =0, \Upsilon>0,
\Lambda>0$ we have the solution

\[ a(\tau) = a_0 \cosh^{1/3}(\sqrt{3\Lambda} \tau)  \]

Now, in such a time-symmetric universe the horizon is, if not
infinite, at least big enough to solve the cosmological horizon
problem (cf. \cite{Primack}) without inflation.  Because the flatness
of the universe does not need explanation too, there is no necessity
for inflation theory.  This qualitative property remains valid for
arbitrary small values $\Upsilon >0$.  The evidence for a hot state of the
universe gives upper bounds for $\Upsilon $.

\subsection{$\Upsilon >0$ stops gravitational collapse before horizon formation}

Now, cosmological observation gives upper limits for $\Xi , \Upsilon $.  For
computations in the solar system, it is possible to use the ``GR
approximation'' $\Upsilon ,\Xi \to 0$. But for strong gravitational fields
they may become important again.  Let's describe how to detect the
domain of application of this GR approximation.

Let's assume we have a GR solution. First, we have to find the correct
Galilean coordinates.  For this purpose we have to define appropriate
initial and boundary conditions for these coordinates. They may be
obtained from gluing with the global universe solution, or from
symmetry considerations.  For example, for a spherically symmetric
stable star we use harmonic coordinates which make the solution
spherically symmetric and stable:

\[ ds^2 = (1-{mm'\over r})
     	    	({r-m\over r+m}dt^2-{r+m\over r-m}dr^2)
       	- (r+m)^2 d\Omega^2 \]

(the function m(r) with $0<m<r, m'>0$ defines the mass inside the
sphere in appropriate units).  For a collapsing star, these
coordinates may be used as initial values.  Once we have found the
preferred Galilean coordinates, we have to prove if $g^{\mu\nu}$
remains small enough. Else, the GR approximation becomes invalid.

For example, for the Schwarzschild solution this happens near the
horizon.  The ansatz $m(r)=(1-\Delta)r$ defines a stable solution for
$p=\varepsilon$:

\begin{eqnarray*}
ds^2 &=& \Delta^2dt^2 - (2-\Delta)^2(dr^2+r^2d\Omega^2) \\
0    &=& -\Upsilon \Delta^{-2} +3\Xi(2-\Delta)^{-2}+\Lambda+\varepsilon\\
0    &=& +\Upsilon \Delta^{-2} + \Xi(2-\Delta)^{-2}+\Lambda-\varepsilon
\end{eqnarray*}

Even if $\Upsilon,\Xi,\Lambda\approx 0$, for $\Delta\ll 1$ we can
ignore only the terms with $\Xi, \Lambda$, but not the
$\Upsilon$-term. We obtain a time-independent solution
$\varepsilon=\Upsilon\Delta^{-2}>0$ for the inner part of a star, with
time dilation $\Delta^{-1}=\sqrt{\varepsilon/\Upsilon}$.
Once no horizon exists, the old notion ``frozen star'' seems more
appropriate than ``black hole''.  Frozen stars remain visible, but
highly redshifted for small $\Upsilon$.

If we interpret for example quasars as frozen stars, this leads to a
relation between redshift and mass: $\Delta\sim M$.

\subsection{The cutoff length in quantum gravity}

Quantization of a condensed matter theory in a classical Newtonian
framework is essentially simpler compared with GR quantization.  The
preferred Newtonian framework avoids most conceptual problems (problem
of time \cite{Isham}, topological foam, information loss problem),
allows to define uniquely local energy and momentum density for the
gravitational field as well as the Fock space and vacuum state in
semi-classical theory.  

What remains are the ultraviolet problems.  But they may be cured by
explicit, physical regularization if we accept an ``atomic
hypothesis'' in our condensed matter interpretation.  Unlike in
renormalized QFT, the relationship between bare and renormalized
parameters obtains a physical meaning.

Similar ideas are quite old and in some aspects commonly accepted
among particle physicists \cite{Jegerlehner}.  Usually it is expected
that the critical cutoff length is of order of the Planck length $a_P
\approx 10^{-33}cm$ \cite{Jegerlehner},\cite{Volovik}.  But an atomic
hypothesis for our condensed matter interpretation predicts a
different cutoff: Once we interpret $\rho$ as the number of ``atoms''
per volume, we obtain the prediction

\[ \rho(x) V_{cutoff} = \mbox{cons}. \]

Considering this prediction for the homogeneous universe, we find that
the cutoff length seems to expand together with the universe. More
accurate, our rulers shrink compared with the cutoff length.


\begin{thebibliography}{99}

\bibitem{Anandan}
J.~Anandan, Phys. Rev. D53, nr. 2, p.779-786, 1996

\bibitem{Aspect}
A. Aspect et.al., Phys. Rev.Lett. 49,
1801-1807, 1982

\bibitem{Bell}
J.S.~Bell, Speakable and unspeakable in quantum mechanics, Cambridge
University Press, Cambridge, 1987

\bibitem{Bohm}
D.~Bohm, Phys. Rev. 85, 166-193, 1952

\bibitem{EPR}
A.~Einstein, B.~Podolsky, N.~Rosen, Phys. Rev. 47, 777-780,
1935

\bibitem{Fock}
V.~Fock, Theorie von Raum, Zeit und Gravitation, Akademie-Verlag
Berlin, 1960

\bibitem{Isham}
C.~Isham, gr-qc/9210011

\bibitem{Jegerlehner}
F.~Jegerlehner, hep-th/9803021

\bibitem{Kuchar}
K.V.~Kuchar, C.G.~Torre, Physical
Review D, vol. 44, nr. 10, pp. 3116-3123, 1991

\bibitem{Logunov}
A.A.~Logunov et.al., Theor.Math.Phys. 69 Nr 3, 1986

A.A.~Logunov, Theor.Math.Phys. 70 Nr 1, 1987

\bibitem{Primack}
J.R.~Primack, astro-ph/9707285

\bibitem{Schmelzer}
I.~Schmelzer, gr-qc/9605013

\bibitem{Volovik} 
G.E.~Volovik, cond-mat/9806010

\end{thebibliography}
\end{document}